\documentclass[preprintnumbers, prd, twocolumn, showpacs, floatfix, preprintnumbers, letterpaper, nofootinbib, amsmath, amssymb, superscriptaddress]
{revtex4-1}
\usepackage{graphicx}
\usepackage{epsfig}
\usepackage{bm}
\usepackage{amsfonts}

\usepackage{color}

\newbox\pippobox

\def\be{\begin{equation}}
\def\ee{\end{equation}}
\def\ba{\begin{eqnarray} }
\def\ea{\end{eqnarray}}

\newcommand {\lla} {\ {\raise-.5ex\hbox{$\buildrel<\over\sim$}}\ }

\usepackage[T1]{fontenc}
\usepackage[latin1]{inputenc}
\usepackage{graphicx}
\usepackage[english]{babel}
\usepackage{amsmath}
\usepackage{amssymb}
\usepackage{amsfonts}

\def\be{\begin{equation}}
\def\ee{\end{equation}}
\def\bea{\begin{eqnarray}}
\def\eea{\end{eqnarray}}

\begin{document}

\title{Charged black holes in nonlinear massive gravity}

\author{Yi-Fu Cai \footnote{Email: yifucai@physics.mcgill.ca}}
\affiliation{Department of Physics, McGill University, Montr\'eal, QC, H3A 2T8, Canada}
\affiliation{Department of Physics, Arizona State University, Tempe, AZ 85287, USA}

\author{Damien A. Easson \footnote{Email: easson@asu.edu}}
\affiliation{Department of Physics, Arizona State University, Tempe, AZ 85287, USA}

\author{Caixia Gao \footnote{Email: cgao1@go.olemiss.edu}}
\affiliation{Department of Physics and Astronomy, University of Mississippi, University, MS 38677, USA}

\author{Emmanuel N. Saridakis \footnote{Email: Emmanuel$_-$Saridakis@baylor.edu}}
\affiliation{Physics Division, National Technical University of Athens, 15780 Zografou Campus, Athens, Greece}
\affiliation{CASPER, Physics Department, Baylor University, Waco, TX 76798-7310, USA}


\begin{abstract}
We find static charged black hole solutions in  nonlinear massive
gravity. In the parameter space of two gravitational potential parameters $(\alpha, \beta)$ we
show that below the Compton wavelength the black hole solutions reduce 
to that of Reissner-Nordstr\"om via the Vainshtein
mechanism in the weak field limit. In the simplest case with
$\alpha=\beta=0$ the solution exhibits the vDVZ discontinuity but
ordinary General Relativity is recovered deep inside the horizon due to the existence of
electric charge. For  $\alpha\neq0$ and $\beta=0$,
the post-Newtonian parameter of the charged black hole evolves to that of
General Relativity via the Vainshtein mechanism within a macroscopic
distance; however, a logarithmic correction to the metric factor of the time
coordinate is obtained. When $\alpha$ and
$\beta$ are both nonzero, there exist two branches of solutions depending
on the positivity of $\beta$. When $\beta<0$, the strong coupling of the
scalar graviton weakens gravity at distances smaller than the Vainshtein
radius. However, when $\beta>0$ the metric factors  exhibit only small
corrections compared to the solutions obtained in General Relativity, and
under a particular choice of $\beta=\alpha^2/6$ the standard
Reissner-Nordstr\"om-de Sitter solution is recovered.
\end{abstract}

\maketitle

\section{Introduction}

The question on whether there exits a consistent covariant theory for
massive gravity, where the graviton acquires a mass and leads to a
modification of General Relativity, was initiated by Fierz and Pauli (FP) \cite{Fierz:1939ix}. It was observed that at quadratic
order the FP mass term is the only ghost-free term describing a gravity
theory with five degrees of freedom  \cite{VanNieuwenhuizen:1973fi}.
However, it is not possible to recover linearized Einstein gravity in the limit
of vanishing graviton mass, due to the existence of the van
Dam-Veltman-Zakharov (vDVZ) discontinuity arising from the coupling between the
longitude mode of the graviton and the trace of the energy momentum tensor
\cite{vanDam:1970vg, Zakharov:1970cc}. It was later observed that this
troublesome mode could be suppressed at macroscopic length scales due to
a nonlinear effect: the so-called Vainshtein mechanism
\cite{Vainshtein:1972sx}. However, these nonlinear terms, which are
responsible for the suppression of vDVZ discontinuity, lead inevitably to
the existence of the Boulware-Deser (BD) ghost  \cite{Boulware:1973my}, making the theory unstable  \cite{ArkaniHamed:2002sp, Creminelli:2005qk,
Deffayet:2005ys, Gabadadze:2003jq}.

Although for many years it was believed that the theory of massive gravity
always contains BD ghosts, a family of its nonlinear extension was recently
constructed by de Rham, Gabadadze and Tolley (dRGT)  \cite{deRham:2010ik,
deRham:2010kj}. This is a two parameter family of nonlinear generalization
of the FP theory, where the BD ghosts are removed in the decoupling limit
to all orders in perturbation theory through a systematic construction of a
covariant nonlinear action \cite{Hassan:2011vm, Hassan:2011hr,
deRham:2011rn, deRham:2011qq} (see  \cite{Hinterbichler:2011tt} for a
review). As a consequence, the theoretical and phenomenological advantages
of the dRGT theory led to a wide investigation in the literature. For example, cosmological implications of the
dRGT theory are  discussed in \cite{D'Amico:2011jj,
Gumrukcuoglu:2011ew, Gumrukcuoglu:2011zh, DeFelice:2012mx,
Gumrukcuoglu:2012aa, Koyama:2011wx, Comelli:2011zm, Crisostomi:2012db,
Cardone:2012qq, Gratia:2012wt, Kobayashi:2012fz, D'Amico:2012pi,
Fasiello:2012rw, D'Amico:2012zv, Langlois:2012hk, Gong:2012yv,
Huang:2012pe, Saridakis:2012jy, Cai:2012ag, Chiang:2012vh}; black holes and
spherically symmetric solutions were analyzed in
\cite{Koyama:2011xz, Koyama:2011yg, Sbisa:2012zk, Nieuwenhuizen:2011sq,
Gruzinov:2011mm, Comelli:2011wq, Berezhiani:2011mt, Sjors:2011iv,
Brihaye:2011aa}; and the theory's connections to bi-metric gravity models were
studied in   \cite{Damour:2002wu, Hassan:2011tf, Hassan:2011zd,
vonStrauss:2011mq, Hassan:2011ea, Hassan:2012wr, Volkov:2011an,
Volkov:2012wp, Paulos:2012xe, Hinterbichler:2012cn, Baccetti:2012bk,
Baccetti:2012re, Baccetti:2012ge, Berg:2012kn}.

Among these phenomenological studies, is the search for observationally 
suitable spherically symmetry solutions. Theories of massive gravity can be strongly constrained due to the vDVZ
discontinuity appearing in the post-Newtonian parameters. Recently, a class
of black hole solutions in the theory of ``ghost-free'' massive gravity was 
analyzed by Koyama, Niz and Tasinato (KNT)
\cite{Koyama:2011xz, Koyama:2011yg}. Their result shows that the behavior
of linearized solutions in General Relativity can only be reproduced below
the Vainshtein distance in a certain region of parameter space. An exact
Schwarzschild-de Sitter (SdS) solution was found for a group of specially selected parameters
\cite{Nieuwenhuizen:2011sq}.

In this paper we present a family of static,
electrically charged black hole
solutions in the theory of ghost-free massive gravity. 
The solutions posses a Vainshtein
radius, below which the linearized solutions of Einstein
gravity are approximately recovered in the weak charge limit.
In massive gravity the longitudinal mode of gravitons is
strongly coupled to the trace of energy momentum tensor and the existence of
an electric charge can strongly affect the behavior of this mode. 

The paper is organized as follows. In Section \ref{model}, we review the
model of nonlinear massive gravity and present the equations of
motion for gravitational fields in a spherically symmetric background. In
section \ref{sphericalsol} we investigate in detail a stellar
background with a static electric field.
In
section \ref{Vainshteinmech} we analyze the charged black
hole solutions both analytically and numerically, and we show that the
Vainshtein effect can be made manifest in a certain parameter space in the weak
field limit. Finally, section \ref{Conclusions} summarizes our results.

\section{Ghost-free massive gravity}
\label{model}

Massive gravity has an effective field theory description given by
Einstein gravity plus the covariant FP mass term. For the dRGT model
Lagrangian the potentially pathological term can be absorbed by total derivative terms, leading
to  equations of motion that are at most second order in
time derivatives \cite{deRham:2010kj}.

\subsection{The dRGT action}

The gravitational action is:
\begin{eqnarray}\label{action}
 S = \int d^4x \sqrt{-g} \frac{1}{16\pi G} \bigg[ R +m^2 {\cal U}(g, \phi^a)
\bigg] ~,
\end{eqnarray}
where $R$ is the Ricci scalar, and ${\cal U}$ is a potential for the graviton which modifies the
gravitational sector. Specifically, ${\cal U}$ is given by
\begin{eqnarray}\label{potential}
 {\cal U}(g, \phi^a) = {\cal U}_2 + \alpha_3{\cal U}_3 +\alpha_4{\cal U}_4
~,
\end{eqnarray}
in which $\alpha_3$ and $\alpha_4$ are dimensionless parameters.
Moreover, ${\cal U}_2$, ${\cal U}_3$ and ${\cal U}_4$ are defined as
\begin{eqnarray}
 {\cal U}_2&\equiv&[{\cal K}]^2-[{\cal K}^2] ~,\\
 {\cal U}_3&\equiv&[{\cal K}]^3-3[{\cal K}][{\cal K}^2]+2[{\cal K}^3] ~,\\
 {\cal U}_4&\equiv&[{\cal K}]^4-6[{\cal K}]^2[{\cal K}^2]+8[{\cal K}][{\cal
K}^3]+3[{\cal K}^2]^2-6[{\cal K}^4] ,\ \
\end{eqnarray}
with
\begin{eqnarray}
 {\cal K}^\mu_\nu =
\delta^\mu_\nu-\sqrt{g^{\mu\sigma}\eta_{ab}
\partial_\sigma\phi^a\partial_\nu\phi^b}~,
\end{eqnarray}
where the rectangular brackets denote the traces,
namely $[{\cal K}]={\cal K}^\mu_\mu$. Finally, in the above relation  the
four-form fields $\phi^a$ are the St\"uckelberg scalars introduced to
restore general covariance \cite{ArkaniHamed:2002sp}.

\subsection{Generalized Einstein Equations}

For convenience  we choose the unitary gauge $\phi^a=x^\mu\delta^a_\mu$ and
thus the tensor $g_{\mu\nu}$ is the observable describing the five degrees
of freedom of the massive graviton. In addition, we regroup the two
parameters $\alpha_3$ and $\alpha_4$ of the graviton potential \eqref{potential} introducing two new parameters, $\alpha$ and
$\beta$, as
\begin{eqnarray}\label{alphabeta}
 \alpha_3 = \frac{\alpha-1}{3}~,~~\alpha_4 =
\frac{\beta}{2}+\frac{1-\alpha}{12}~,
\end{eqnarray}
to simplify the background equations of motion.

Varying the action with respect to $g_{\mu\nu}$  leads to the
modified Einstein equations:
\begin{eqnarray}\label{EoM}
 G_{\mu\nu} +m^2X_{\mu\nu} = 8\pi G T_{\mu\nu}~,
\end{eqnarray}
where $X_{\mu\nu}$ arises from the graviton potential \footnote{Note
that the Einstein equation shown in  \cite{Berezhiani:2011mt}
contains a typo of an extra $1/2$ factor, however the
equations of motion which give rise to solutions were based
on the method of varying the action with respect to the metric factor
directly. }
\begin{eqnarray}
 X_{\mu\nu} &=& {\cal K}_ {\mu\nu} -{\cal K}g_ {\mu\nu} \nonumber\\
 && -\alpha\left\{{\cal K}^2_{\mu\nu}-{\cal K}{\cal K}_{\mu\nu} +\frac{[{\cal K}]^2-[{\cal K}^2]}{2}g_{\mu\nu}\right\} \nonumber\\
 && +6\beta\left\{ {\cal K}^3_{\mu\nu} -{\cal K}{\cal K}^2_{\mu\nu} +\frac{1}{2}{\cal K}_{\mu\nu}\left\{[{\cal K}]^2 -[{\cal K}^2]\right\}\right\} \nonumber\\
 &&-\beta g_{\mu\nu}\left\{[{\cal K}]^3 -3[{\cal K}][{\cal K}^2] +2[{\cal K}^3]\right\}
~.
\end{eqnarray}
In addition to the generalized Einstein equations, the Bianchi identities lead to
the constraint:
\begin{eqnarray}\label{BiEoM}
 \nabla^\mu X_{\mu\nu} = 0~.
\end{eqnarray}

\section{General analysis on spherically symmetric charged background}
\label{sphericalsol}

Having derived the equations of motion we now study the
dynamics of a gravitational system described by massive gravity under a
fixed background symmetry. Following \cite{Koyama:2011xz}, 
we consider the most general form of
the metric respecting spherical symmetry,
\begin{eqnarray}\label{metric}
 ds^2 = -N^2(r)dt^2 +\frac{dr^2}{F^2(r)} +2D(r)dtdr
+\frac{r^2d\Omega_2^2}{H^2(r)}~,
\end{eqnarray}
where $d\Omega_2^2 =d\theta^2 +\sin^2\theta d\varphi^2$.

Notice that  the identity $g_{0r}R_{00}-g_{00}R_{0r}=0$ yields an algebraic
constraint and correspondingly leads to two solution branches: either
$D(r)=0$, or the effective energy density is proportional to $g_{tt}$.
The latter case always yields a constant metric factor $H(r)$ with value depending on specific backgrounds
\cite{Berezhiani:2011mt}. In the following we will focus on the first case
$D(r)=0$, corresponding to a diagonal metric.

\subsection{Charged spherical symmetric background in General Relativity}

We consider a generic Maxwell field $F_{\mu\nu}$ in curved spacetime, with
standard Lagrangian. The Maxwell equations are
\begin{eqnarray}
 \partial_\mu(\sqrt{-g} F^{\mu\nu}) = -\sqrt{-g} J^\nu~,
\end{eqnarray}
where $J^\nu$ is the current density. For a static
electric charge $Q$ in the gravitational system, the components of the Maxwell
field are:

\begin{eqnarray}
 E_r=F_{0r}=E(r)~,~~E_{\theta}=E_{\varphi}=0~,~~\vec{B}=0~.
\end{eqnarray}

With vanishing source term, the inhomogeneous Maxwell law gives
\begin{eqnarray}
 \partial_r(\sqrt{-g}F^{0r})=0~,
\end{eqnarray}
yielding  the   solution:
\begin{eqnarray}\label{Er}
 E(r) = \frac{QNH}{4\pi F r^2}~,
\end{eqnarray}
where $Q$ is an integration constant which is typically interpreted as the
electric charge. The factors $N$, $F$, and $H$ are the fields introduced in
the background metric \eqref{metric}.
Varying the electromagnetic Lagrangian with respect to the
metric gives  the energy momentum tensor:
\begin{eqnarray}\label{Tmunu}
 T_{\mu\nu} =
F_{\mu\sigma}F^\sigma_\nu-\frac{F^{\mu\nu}F_{\mu\nu}}{4}g_{\mu\nu}~.
\end{eqnarray}

It is well known that  a static and spherically symmetric solution
under standard General Relativity is described by the
Reissner-Nordstr\"om (RN) solution:
\begin{eqnarray}\label{RNsolution}
 ds^2 =
-{\left(1+\frac{r_Q^2}{r^2}-\frac{r_S}{r}\right)}dt^2+\frac{dr^2}{1+\frac{
r_Q^2 } { r^2 } -\frac{r_S}{r}} +r^2d\Omega^2~,\ \
\end{eqnarray}
with
\begin{eqnarray}\label{r_Q}
 r_Q\equiv \sqrt{\frac{GQ^2}{4\pi}}~,~~r_S\equiv 2GM~
\end{eqnarray}
being respectively the length scale associated with the electric charge $Q$
and the Schwarzschild radius determined by the mass of the spherical object
$M$. The above solution exhibits a singularity at
$r=0$, in which the metric coefficient becomes zero and invariants like the
Ricci and Kretschmann scalars diverge, however it is usually shielded by a
horizon at $r=\left(r_S+\sqrt{r_S^2-4r_Q^2}\right)/2$, in which the metric
coefficients become zero in the above coordinates but the invariants remain
finite, thus satisfying the cosmic censorship and no-hair conjectures. 
Note however that negative mass or highly charged solutions
do not have a horizon, and thus a naked singularity appears.

\subsection{Equations of motion for spherically symmetric charged background}

Going beyond the above General Relativity solution, in the case of nonlinear
massive gravity, we must consider the effects of the graviton potential. Thus one can
combine the generalized Einstein equations \eqref{EoM}, the solution to the
Maxwell field \eqref{Er}, and the energy momentum tensor \eqref{Tmunu}, and
obtain three main equations of motion.

We consider the spherically symmetric
metric Ansatz \eqref{metric} with $D(r)=0$. The background equations
of motion are derived from the generalized Einstein
equations \eqref{EoM}. After some algebra, we extract the
equations of motion:
For the $``00"$ component of generalized Einstein equation:
\begin{eqnarray}\label{EoMtt}
 &&\frac{GQ^2H^6}{4\pi r^2} = (1+m^2r^2)H^4 +2m^2r^2(F-3)H^3 \nonumber\\
 && -H^2[2r(F\dot{F}-3m^2r^2)+3m^2r^2F+F^2] \nonumber\\
 && +2rFH[F(r\ddot{H}+3\dot{H})+r\dot{F}\dot{H}] -5r^2F^2\dot{H}^2
\nonumber\\
 && +m^2r^2(H-1)H^2 [ 2H(1-\alpha+3\beta) -6\beta \nonumber\\
 && +F(1 -\alpha +6\beta -3H(1 -\alpha +2\beta)) ]  ~.
\end{eqnarray}
The $``rr"$ component of generalized Einstein equation takes the form:
\begin{eqnarray}\label{EoMrr}
 &&\frac{GQ^2NH^6}{4\pi r^2} = (1+m^2r^2)NH^4 +2m^2r^2(1-3N)H^3 \nonumber\\
 && +H^2[N(6m^2r^2-F^2)-r(2F^2\dot{N}+3m^2r^2)] \nonumber\\
 && +2rF^2H\dot{H}(r\dot{N}+N) -r^2F^2N\dot{H}^2 \nonumber\\
 && +m^2r^2(H-1)H^2 [ 1-\alpha+6\beta(1-N) \nonumber\\
 && -H(3(1-\alpha)-2N(1-\alpha+3\beta) +6\beta) ] ~.
\end{eqnarray}
Further, the generalized Einstein equation along the solid angle is given
by,
\begin{eqnarray}\label{EoMth}
 &&\frac{GQ^2NH^6}{4\pi r^3} = m^2rH^3[(3-F)N-1] +2rF^2N\dot{H}^2
\nonumber\\
 && -F H[rN\dot{F}\dot{H}+F(rN\ddot{H}+r\dot{N}\dot{H}+2N\dot{H})]
\nonumber\\
 && +H^2[rF\dot{F}\dot{N}+NF\dot{F}+F^2(r\ddot{N}+\dot{N}) \nonumber\\
 && +m^2rF(3N-1)-6m^2rN+3m^2r ] \nonumber\\
 && +m^2rH^2 \{ (1-\alpha)[ 4N -3 +H(2-3N) \nonumber\\
 && +F(2-3N+H(2N-1))] \nonumber\\
 &&  + 2(F-1)(H-1)(N-1)(2-2\alpha+3\beta) \} ~.
\end{eqnarray}

The equations of motion may also be obtained by varying the
action with respect to the metric fields $N$, $F$ and $H$, respectively.
In addition, there is another constraint equation from the Bianchi
identity \eqref{BiEoM}:
\begin{eqnarray}\label{EoMBi}
 &&0 = \frac{1}{rNH} \left\{ F\bigg\{ H[2-3r\dot{N}-2N(3+r\dot{H})]
\right.\nonumber\\
 && +2(3N-1)r\dot{H} +2H^2(r\dot{N}+N) \bigg\} - 2H^2[1+N(H-3)] \nonumber\\
 && + (1-\alpha) \bigg\{ 2H^2 [2-3N+H(2N-1)] \nonumber\\
 && + F[ r\dot{N}H^3 +2r\dot{H}(2-3N) +H^2(2-4N-4r\dot{N}) \nonumber\\
 &&  +H(6N-4-2r\dot{H}+3r\dot{N}+4r\dot{H}N) ] \bigg\} \nonumber\\
 &&  +2(2\alpha-3\beta-2) (H-1) \bigg\{2H^2(N-1) \nonumber\\
 &&   \left. +F[r\dot{N}H^2 +2r\dot{H}(N-1) -H(2N-2 +r\dot{N})]\bigg\}
\right\}~,
\end{eqnarray}
where in the above equations the dot denotes a derivative with respect to
the radial coordinate $r$.

Note that when we take $Q=0$ the above
equations of motion reduce to the case of the strong interactive
solar system governed by nonlinear massive gravity, discussed in \cite{Koyama:2011yg}. 
Our expressions are in agreement with theirs, expect that the radial metric
factor is $F^2$ in our case but becomes $F$ in their convention. The
relations between the parameter spaces $(\alpha_3, \alpha_4)$ and $(\alpha,
\beta)$ are already given in  \eqref{alphabeta}.
Equation \eqref{EoMBi} obtained from the Bianchi identity constraint is not an independent equation 
after we choose $D(r)=0$.

\subsection{The linearized treatment of nonlinear massive gravity}

Following the idea developed by KNT  \cite{Koyama:2011xz, Koyama:2011yg},
we study the solutions to such a gravitational system in the weak field
limit. We can expand the metric factors around a Minkowski background as
\begin{eqnarray}
 N(r)&=&1+n(r)~,~\nonumber\\
 F(r)&=&1+f(r)~,~\nonumber\\
 H(r)&=&1+h(r)~,
\end{eqnarray}
and then investigate the linear perturbations. However, we need to be aware
of the fact that the factors $n$ and $f$ can be treated as linear
perturbations as in General Relativity, while $h$ could, in principle, take large values
since this factor corresponds to the strong interactive nature of the
scalar mode of graviton in solar system. Therefore, we need to keep higher
orders in $h$ and truncate equations of motion to leading order of $n$ and
$f$. We demonstrate this behavior both in analytical and numerical
calculations in the following.

Before expanding the background equations perturbatively, we 
rescale the radial coordinate by introducing a new metric variable
\begin{eqnarray}
 \rho \equiv \frac{r}{H}~,
\end{eqnarray}
and correspondingly introduce a new metric factor
\begin{eqnarray}
 1+\tilde{f} = \frac{1+f}{1+h+\rho h'}~,
\end{eqnarray}
where the prime denotes a derivative with respect to $\rho$. As a
consequence, the linearized metric can be expressed as
\begin{eqnarray}\label{metricpert}
 ds^2 = -[1+2n(\rho)]dt^2+[1-2\tilde{f}(\rho)]d\rho^2+\rho^2d\Omega^2
\end{eqnarray}
which is asymptotic to Minkowski background when $n$ and $\tilde{f}$ become
negligible.


Apart from the usual curvature invariants, nonlinear massive gravity
presents a new basic invariant incorporating both the metric and the
St\"uckelberg scalars, namely $I^{ab}\equiv g^{\mu\nu}
\partial_\mu \phi^a \partial_\nu \phi^b$. Under unitary gauge it seems
that this invariant encounters a divergence on the event horizon if one
takes the Minkowskian asymptotic at large scales \cite{Deffayet:2011rh},
that is, one obtains singularities in the place where General Relativity
had simple horizons. The authors of \cite{Berezhiani:2011mt} argue that
a black hole solution to nonlinear massive gravity might be viable only
when the invariant $I^{ab}$ is non-singular. Though its divergence does not
bring any manifest problem to observable variables at background level, it
may be a problem for fluctuations passing through the horizon. In the
present paper, we focus on the dynamics of background solutions outside the
horizon and thus we do not address this issue. 

\subsection{The post-Newtonian parameter and the vDVZ discontinuity}

To study the nontrivial effects of massive gravity we consider
the metric factor $h$ as a perturbation mode.
By expanding the equations of motion \eqref{EoMtt},
\eqref{EoMrr} and \eqref{EoMBi} presented in the Appendix, consistent
with the linearized background \eqref{metricpert}, we get
\begin{eqnarray}
\label{EoMperth1} & (2+m^2\rho^2)\tilde{f} +2\rho\tilde{f}'
+m^2\rho^2(3h+\rho h') = -\frac{GQ^2}{4\pi\rho^2}~, \\
\label{EoMperth2} & 2\rho n' +2\tilde{f} +m^2\rho^2(2h-n) =
-\frac{GQ^2}{4\pi\rho^2}\left(1+n\right)~, \\
\label{EoMperth3} & \rho n' +2\tilde{f} = 0~,
\end{eqnarray}
up to leading order.

Under the above approximation the metric
factors are:
\begin{eqnarray}\label{na}
 n(\rho) &\simeq& -\frac{4GM e^{-m\rho}}{3\rho} +\frac{GQ^2}{8\pi\rho^2}
\nonumber\\
  && +\frac{GmQ^2}{16\pi\rho}  \left[e^{m\rho}{\rm
Ei}(-m\rho)-e^{-m\rho}{\rm Ei}(m\rho)\right]~, \ \
\end{eqnarray}
\begin{eqnarray}\label{fa}
 \tilde{f}(\rho) &\simeq& -\frac{2GM e^{-m\rho} (1+m\rho)}{3\rho}
+\frac{GQ^2}{8\pi\rho^2} \nonumber\\
 && +\frac{GmQ^2}{32\pi\rho} \times \left[(1-m\rho)e^{m\rho}{\rm
Ei}(-m\rho)\right. \nonumber\\
 &&\left. ~~~~ -(1+m\rho)e^{-m\rho}{\rm Ei}(m\rho)\right]~,
\end{eqnarray}
and
\begin{eqnarray}\label{ha}
 h(\rho) &\simeq& -\frac{2GMe^{-m\rho}}{3m^2\rho^3}(1+m\rho+m^2\rho^2)
+\frac{GQ^2}{16\pi\rho^2} \nonumber\\
 && +\frac{GQ^2}{32\pi m\rho^3} \times
\left[(1-m\rho+m^2\rho^2)e^{m\rho}{\rm Ei}(-m\rho)\right. \nonumber\\
 &&\left. ~~~~ -(1+m\rho+m^2\rho^2)e^{-m\rho}{\rm Ei}(m\rho)\right]~,
\end{eqnarray}
where $``{\rm Ei}"$ is the exponential integral function defined by ${\rm
Ei}(x)\equiv \int_{-\infty}^x e^{t}~ d\ln{t}$.

Let us now examine the post-Newtonian parameters for nonlinear massive
gravity. The  post-Newtonian parameters are strongly constrained by
solar system observations (see  \cite{Will:2005va} for a detailed
introduction  and a comprehensive review),
and therefore may be used to place strong constraints on a given theory.  The first
post-Newtonian parameter $\gamma$  is defined
as the ratio of $\tilde{f}$ and $n$:
\begin{eqnarray}\label{gamma}
 \gamma\equiv \frac{\tilde{f}}{n}~,
\end{eqnarray}
in the weak field limit. In the case of General Relativity $\gamma=1$, as
we can immediately read from the RN solution \eqref{RNsolution}. However,
in the regime of length scales much smaller than the Compton wavelength
($\rho_m=1/m$) in nonlinear massive gravity, this parameter can be
approximately expressed as
\begin{eqnarray}\label{gamma_a}
 \gamma \simeq \frac{32\pi M\rho(1+m\rho)-6Q^2e^{m\rho}}{64\pi
M\rho-6Q^2e^{m\rho}}~,
\end{eqnarray}
where we have made the reasonable assumption that the solar-system mass $M$
is much larger than graviton mass.

It is straightforward to see that when the solar system
does not carry an electric charge ($Q=0$), we obtain
$\gamma\simeq(1+m\rho)/2$, and thus in the massless limit it reduces to
$\gamma=1/2$, which is in stark disagreement with the value in General Relativity. This
behavior exactly manifests the vDVZ discontinuity, and thus we face the
difficulty of explaining solar-system observations at present.
However, when the contribution of $Q$ is taken into account, the
post-Newtonian parameter could approach to $1$ when $\rho$ is much smaller
than the length scale
\begin{eqnarray}
 \rho_G=\frac{3Q^2}{32\pi M}~.
\end{eqnarray}
This result implies that the vDVZ discontinuity  exists in a charged
solar system governed by nonlinear massive gravity, but the explicit
behavior is different from a electro-neutral one.

Finally, from the solutions \eqref{na}, \eqref{fa} and \eqref{ha} we
 find that the weak-field limit is a good approximation at large
distances. However, when the radial coordinate $\rho$ decreases below a
critical value
\begin{eqnarray}
 \rho_V \equiv \left(\frac{GM}{m^2}\right)^{1/3}~,
\end{eqnarray}
the absolute value of the factor $h$ will increase exponentially and become
much larger than unity. This critical radius is the so-called Vainshtein
radius. We perform a detailed analysis on the perturbation equations
by keeping all nonlinear order terms of $h$ in the next section.

\section{Charged Black holes and Vainshtein mechanism}
\label{Vainshteinmech}

Due to  the famous Vainshtein mechanism
\cite{Vainshtein:1972sx}  the scalar degree of freedom in massive gravity
becomes strongly coupled in the limit of small graviton mass, and thus the
linearized treatment performed in equations \eqref{EoMperth1},
\eqref{EoMperth2} and \eqref{EoMperth3} is no longer reliable. This is
observed by following the evolution of metric factor $h$ below the
Vainshtein radius, where the absolute value of $h$ becomes much greater
than unity. Consequently, although we can keep treating $n$ and $\tilde{f}$
as small perturbations in this regime, higher order terms in $h$ should
be taken into account.

\subsection{Perturbation equations with nonlinear corrections}

Keeping leading order in $n$ and $\tilde{f}$, the perturbed equations of
motion including nonlinear corrections of $h$ are given by
\begin{eqnarray}
\label{EoMpertN} && 2\tilde{f} +2\rho\tilde{f}' +m^2\rho^2 \bigg[
(1-2\alpha{h}+6\beta{h}^2)
 [(1+\tilde{f})\rho h' \ \ \ \ \ \ \ \ \nonumber\\
 && ~ \ \ \  +(1+h)\tilde{f}] +3h(1-\alpha{h}+2\beta{h}^2) \bigg]
+\frac{GQ^2}{4\pi\rho^2} = 0,\ \ \ \ \ \\
\label{EoMpertF} && 2\tilde{f} +2\rho{n}' -m^2\rho^2 \bigg[n
-2(1+n+\alpha{n})h \nonumber\\
 && ~ \ \ \ +(\alpha+\alpha{n}+6\beta{n})h^2 \bigg]
+\frac{GQ^2(1+n)}{4\pi\rho^2} = 0~, \\
\label{EoMpertBi} && \rho n'[1-2\alpha{h}+6\beta{h}^2]
+2\tilde{f}[1-\alpha{h}] = 0~,
\end{eqnarray}
of which the first two correspond to the $(00)$ and $(rr)$ components of
the generalized Einstein equations and the last one arises from the
perturbed Bianchi constraint.

We first solve equation  \eqref{EoMpertN} by neglecting all high-order
terms proportional to $m^4$, $G^2M^2$, $m^2GM$, $G^2Q^4$, and $m^2GQ^2$.
Therefore, the metric factor $\tilde{f}$ can be expressed in terms of $h$:
\begin{eqnarray}\label{solf}
 \tilde{f} \simeq \frac{GQ^2}{8\pi\rho^2} -\frac{GM}{\rho}
-\frac{m^2\rho^2}{2}(h-\alpha{h}^2+2\beta{h}^3)~.
\end{eqnarray}
Inserting the expression \eqref{solf} into the perturbation equation
\eqref{EoMpertF}, we obtain the radial derivative of $n$ as a function of
$h$:
\begin{eqnarray}\label{solnp}
 n' \simeq -\frac{GQ^2}{4\pi\rho^3} +\frac{GM}{\rho^2}
-\frac{m^2\rho}{2}(h-2\beta{h}^3)~,
\end{eqnarray}
and thus the metric factor $n$ can be acquired by performing
integration. The key to solving for the metric factors $n$ and
$\tilde{f}$ is to extract the solution for $h$. Therefore, we combine the
expressions \eqref{solf}, \eqref{solnp}, and the perturbed Bianchi
constraint \eqref{EoMpertBi}, and then we derive the polynomial equation:
\begin{eqnarray}\label{solh}
 &&\frac{GM}{\rho}\left(1-6\beta{h}^2\right) -
\frac{GQ^2}{4\pi\rho^2}\left(\alpha{h}-6\beta{h}^2\right) =  \ \ \  \ \ \
\ \ \  \ \ \  \ \ \ \nonumber\\
 &&  \ \ \  \ \ \  m^2\rho^2  \left[ -\frac{3}{2}h +3\alpha{h}^2
-(\alpha^2+4\beta)h^3 +6\beta^2h^5 \right],  \ \ \ \ \
\end{eqnarray}
in which all nonlinear terms of $h$ have been taken into account.

After having obtained equations \eqref{solf}, \eqref{solnp} and
\eqref{solh}, we   are able to calculate the explicit forms of the metric
factors $n$, $\tilde{f}$ and $h$ under different parameter choices of
$(\alpha, \beta)$. In the following subsections, we investigate these
solutions further.

\subsection{Case I: $\alpha=\beta=0$}

Let us first consider the special subclass with $\alpha=\beta=0$. In this
case all higher order terms of $h$ vanish automatically, and thus our task
reduces to solving the linearized equations \eqref{EoMperth1},
\eqref{EoMperth2} and \eqref{EoMperth3}. Therefore, the corresponding
solutions of metric factors are already given in equations  \eqref{na},
\eqref{fa} and \eqref{ha}.

The post-Newtonian parameter takes the value of $1/2$ in the regime of
length scales of interest. Although this parameter increases to $1$ at
very small values of radial coordinate due to the effect of electric
charge, the corresponding length scale is deeply inside the horizon of the
black hole. As a consequence, the standard General Relativity result can not be recovered
in the solar system governed by massive gravity with such a parameter
choice, and therefore this case is already observationally ruled out.

\subsection{Case II: $\alpha\neq 0$ and $\beta=0$}

We now proceed to the case with a vanishing $\beta$ but non-vanishing $\alpha$. 
In principle, one can solve the $h$-equation
\eqref{EoMpertBi} exactly, however the resulting expression is quite
complicated, hiding the underlying physics. Therefore, instead of finding
an exact solution to $h$ we solve the perturbed Bianchi equation
approximately by keeping dominant terms in $h$. Different from the
electro-neutral case of \cite{Koyama:2011yg}, both the   solar-system
mass  $M$ and its electric charge $Q$ contribute to the
L.H.S of equation  \eqref{solh}. Therefore, we need to solve \eqref{solh}
by assuming its L.H.S is dominated by $M$ and $Q$ respectively.

We first consider that the contribution of the electric charge $Q$ becomes
dominant in the L.H.S of \eqref{solh} when $\rho\ll\rho_V$. In this limit
we solve equation  \eqref{solh} and expand $h$ up to order
${\cal{O}}(\rho^2)$:
\begin{eqnarray}\label{hb}
 h \simeq
-\frac{\rho_Q^2}{\alpha^{1/2}\rho^2}+\frac{3}{2\alpha}-\frac{\rho_V^3\rho}{
2\alpha\rho_Q^4}~,
\end{eqnarray}
where we have introduced a new parameter for a critical length scale
\begin{eqnarray}
 \rho_Q \equiv \left(\frac{GQ^2}{4\pi m^2}\right)^{1/4}~.
\end{eqnarray}
One can see that the approximate expression for $h$ is valid only when
$\rho<\rho_Q$. Comparing with the non-charged system analyzed in
\cite{Koyama:2011yg}, we find that the leading term of $h$ in a charged solar
system is proportional to $\rho^{-2}$ instead of $\rho^{-1}$. Moreover, the
suppression scale is $\rho_Q$ rather than the Vainshtein radius $\rho_V$.
However, since for a group of canonical parameters accommodating with
astronomical data  $\rho_V$ is usually much bigger than $\rho_Q$, we
deduce that the contribution of the third term in the R.H.S.
of equation  \eqref{hb} is considerable in a wide regime of length scales.
Then, from equations \eqref{solf} and \eqref{solnp} we can derive
 the expressions for $n$ and $\tilde{f}$ through a  Taylor expansion
\begin{eqnarray}
\label{nb} n &\simeq& \frac{GQ^2}{8\pi\rho^2} -\frac{GM}{\rho}
+\frac{m^2\rho_Q^2}{2\alpha^{1/2}}\ln(m\rho)~,\\
\label{fb} \tilde{f} &\simeq& \frac{GQ^2}{4\pi\rho^2} -\frac{GM}{\rho}
-\frac{m^2\rho_Q^2}{\alpha^{1/2}} +\frac{GM\rho}{2\alpha^{1/2}\rho_Q^2}~,
\end{eqnarray}
up to order ${\cal{O}}(\rho^2)$. By observing \eqref{nb} and \eqref{fb},
when $2GM<\rho<\rho_Q$, the corrections of massive gravity to the General Relativity
results are quite small. When $\rho$ is larger than $\rho_Q$ but smaller
than $\rho_V$, the higher order corrections are suppressed by $\rho_V$ and
the formulae are in agreement with Ref.  \cite{Koyama:2011yg}.

When the radial coordinate $\rho$ evolves to the regime which is close to
$\rho_Q$   but still smaller than $\rho_V$, the main
contribution of the L.H.S of \eqref{solh} comes from the mass term $M$. In
this case, we only keep the leading order in $h$, obtaining
\begin{eqnarray}
\label{hbM} h &\simeq& -\frac{\rho_V}{\alpha^{2/3}\rho}~, \\
\label{nbM} n &\simeq& \frac{GQ^2}{8\pi\rho^2} -\frac{GM}{\rho}
+\frac{GM\rho}{2\alpha^{2/3}\rho_V^2}~, \\
\label{fbM} \tilde{f} &\simeq& \frac{GQ^2}{8\pi\rho^2} -\frac{GM}{\rho}
+\frac{G M}{2\alpha^{1/3}\rho_V} +\frac{GM\rho}{2\alpha^{2/3}\rho_V^2}~.
\end{eqnarray}
Comparing our results with the analysis of
\cite{Koyama:2011yg} we find that the above three solutions are consistent
with those in the solar system without a charge. This implies that there
must exist an intermediate regime along the radial coordinate in which the
behaviors of the metric factors of charged black holes are the same as
those of neutral black holes.

In order to provide a more transparent picture of the dynamics of
the charged solar system described by nonlinear massive gravity with the
above parameter choice ($\alpha\neq0$ and $\beta=0$), we
 numerically evolve the perturbation equations. In particular, without
loss of generality we consider $\alpha=1$, and we choose the graviton mass
$m=10^{-20}M_P$ and the solar-system mass $M=10^{6}M_P$, setting also
the Planck unit $M_P^2=1/G=1$.  Correspondingly, this group of
parameters yield the Compton wavelength $\rho_m=10^{20}$, which is much
larger than the Vainshtein radius $\rho_V=2.15\times10^{15}$. Moreover, we
consider the source term of the matter field to be a weak charge
$Q_W=1.77\times10^{3}$ or a strong charge $Q_S=3.54\times10^{6}$,
respectively.

\begin{figure}[ht]
\begin{center}
\mbox{\epsfig{figure=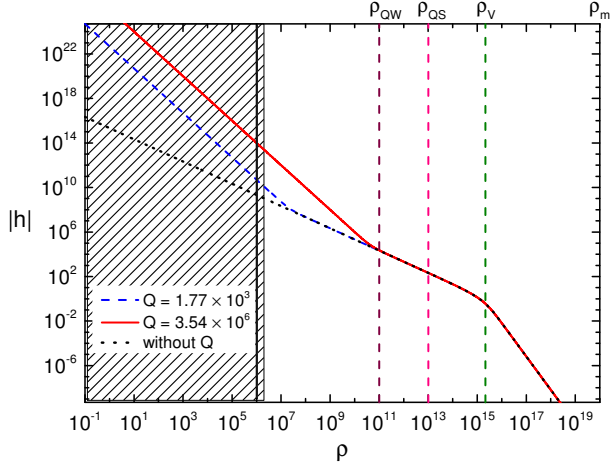,width=7.5cm,angle=-90}} \caption{{\it
Plot of the evolutions of the absolute value of metric factor $h$
as functions of radial coordinate $\rho$ in a charged solar system
described by nonlinear massive gravity. The model parameters are taken as:
$\alpha=1$ and $\beta=0$. In the numerical calculation, we take
$m=10^{-20}$ and $M=10^{6}$. The corresponding Compton wavelength
$\rho_m=10^{20}$ and the Vainshtein radius $\rho_V=2.15\times10^{15}$ are
denoted on the top of the figure. The values of weak and strong electric
charges are provided in the plot. All dimensional parameters are in
Planck units.
}}
\label{Fig:Case2_h}
\end{center}
\end{figure}

In Fig \ref{Fig:Case2_h} we depict the absolute value of the metric factor
$h$ as a function  of the radial coordinate $\rho$. From the above parameter
choices we find that   the suppression scales associated with the
electric charges are given by $\rho_{QW}=2.24\times10^{11}$ (which is
represented as a purple dashed line for the weak charge $Q_W$) and
$\rho_{QS}=10^{13}$ (which is represented as a pink dashed line for the
strong charge $Q_S$). The wide sparse shadow regime denotes the space
between the inner horizon and outer horizon in the case of the weakly
charged solar system, and the narrow dark shadow regime denotes the space
between the two horizons in the case of the strongly charged solar system.

From Fig. \ref{Fig:Case2_h} we that there exist three different
slopes for the metric factor $h$ along the radial coordinate $\rho$ in a
charged solar system. When $\rho$ is greater than the Vainshtein radius
$\rho_V$ but less than the Compton wavelength $\rho_m$, $|h|$ scales
approximately as $\rho^{-3}$  and its value is much smaller than $1$, and
therefore the weak field limit at large distances is valid. When $\rho$
evolves to be smaller than $\rho_V$, $|h|$ becomes much larger than unity
quickly and scales as $\rho^{-1}$. This behavior is in precise agreement
with  \eqref{hbM}. When $\rho$ decreases to a regime which is
much shorter than $\rho_Q$, we observe that the slope of $|h|$ changes
again which gives rise to $|h|\sim\rho^{-2}$. This transition on
$|h|$-slope implies that the contribution of electric charge $Q$ becomes
dominant in determining the dynamics of scalar graviton. Moreover, the
length scale for the slope transition on $h$ depends on the value of
$\rho_Q$ and thus it is determined by the combination of $Q$ and $m$. For a
fixed graviton mass $m$, the value of $\rho_Q$ in a strongly charged solar
system (denoted by the purple dashed line) is much larger   than that
in the case of weak charge (denoted by the pink dashed line).
Correspondingly, the transition of $|h|$-slope in the case of strong
charge occurs at a larger distance as shown in the red solid curve, while
the transition of $|h|$-slope in the case of weak charge takes place at a
smaller distance as shown in the blue dashed curve.

\begin{figure}[ht]
\begin{center}
\mbox{\epsfig{figure=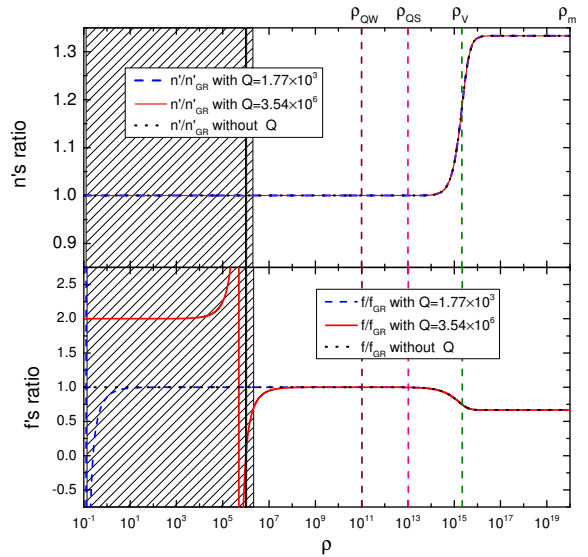,width=7.5cm,angle=-90}} \caption{{\it
Plot of the evolutions of the ratios $n'/{n'}_{GR}$ and
$\tilde{f}/\tilde{f}_{GR}$ as functions of radial coordinate $\rho$ in a
charged solar system described by nonlinear massive gravity. In the numerical
calculation, the parameters are chosen to be the same as those provided in
Fig. \ref{Fig:Case2_h}. The ``$f$'' in the lower panel represents for the
metric factor $\tilde{f}$ in the main text.
}}
\label{Fig:Case2_nf}
\end{center}
\end{figure}

In Fig. \ref{Fig:Case2_nf} we depict  the ratios $n'/{n'}_{GR}$ and
$\tilde{f}/\tilde{f}_{GR}$ as functions of radial coordinate $\rho$.
From the upper graph we observe that the correction to the metric factor $n$ from the electric charge $Q$ is very small, since the three curves (a red solid line representing for strongly
charged case, a blue dashed line representing the weakly charged case, and
a black dotted line representing the non-charged case) almost coincide. 
Additionally, when $\rho$ is outside the Vainshtein radius, the
ratio of $n'$ in nonlinear massive gravity and $n'$ in General Relativity takes
the value of $4/3$. This numerical result is in agreement with the analytic
result obtained in the previous section when one compares the first term in
the R.H.S of equation \eqref{na} and the second term in the R.H.S of
equation \eqref{nb}. However, this ratio tends to $1$ when $\rho$ lies in
the regime of $\rho<\rho_V$ due to the Vainshtein effect. As a consequence,
the metric factor $n$ of nonlinear massive gravity roughly recovers the General Relativity
result inside the Vainshtein radius.

\begin{figure}[ht]
\begin{center}
\mbox{\epsfig{figure=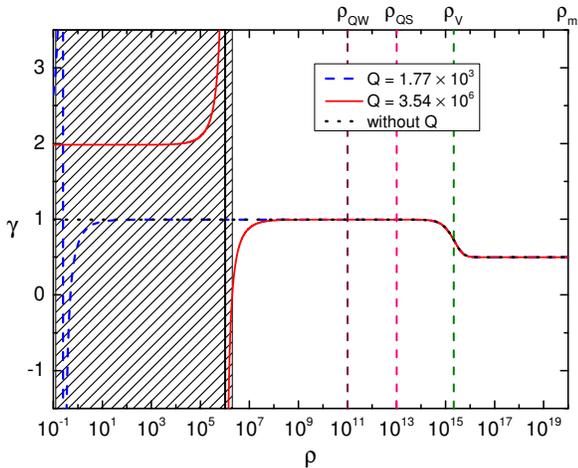,width=7.5cm,angle=-90}} \caption{{\it
Plot of the evolutions of the post-Newtonian parameter $\gamma$ as
functions of radial coordinate $\rho$ in a charged solar system described
by nonlinear massive gravity. In numerical calculation, the parameters are
chosen to be the same as those provided in Fig. \ref{Fig:Case2_h}.
}}
\label{Fig:Case2_gm}
\end{center}
\end{figure}

From the evolutions of $f$ in charged gravitational backgrounds,
depicted in the lower graph of Fig. \ref{Fig:Case2_nf}, we can
see that they are different from the non-charged one when $\rho$ approaches
to the inner horizons. This yields the nontrivial modification to the
post-Newtonian parameter at small length scales as shown in Fig.
\ref{Fig:Case2_gm}. When the radial coordinate is smaller than the inner
horizon of a charged black hole we get $\gamma=2$, which disagrees with
the General Relativity case \footnote{However, one needs
to be aware of the fact that in this regime the perturbative treatment of
$n$ and $\tilde{f}$ is not valid and thus a completely non-perturbative
analysis is required. Such an analysis lies beyond the scope of the
present work.}. From Fig.
\ref{Fig:Case2_gm}, one can see that inside the Vainshtein radius (but
outside the outer horizon) we obtain approximately $\gamma=1$ and thus
such a solar system could conform with observations. When
$\rho$
becomes greater than $\rho_V$, the post-Newtonian parameter $\gamma$
evolves to $1/2$ which is expected by the effect of vDVZ discontinuity as
shown in \eqref{gamma_a}.

\subsection{Case III: $\beta\neq 0$}

The case where both $\alpha$,$\beta$ parameters are non vanishing can be
divided in two subcases: $\beta<0$ and $\beta>0$. Generally, we are unable
to obtain the exact solution of the metric factor $h$ from equation
\eqref{solh}. Therefore, we can only solve the equations of motion
semi-analytically by  keeping the leading order terms and then compare
with numerical computations. However, it is interesting to notice that, in
the case of $\beta>0$ there exists a class of exactly analytic solutions
for a special family of parameter choice $\beta=\alpha^2/6$, which was also
observed in Schwarzschild-like solutions in massive gravity
\cite{Nieuwenhuizen:2011sq, Gruzinov:2011mm,Berezhiani:2011mt}\footnote{An
RN solution was constructed in the de
Sitter (dS) background in \cite{Nieuwenhuizen:2011sq,
Berezhiani:2011mt}. In the present work we solve the exact equations of
motion of a charged solar system in massive gravity and we obtain an exact
form of RN-dS which will be shown in later context. }. In the following we
will analyze these cases in both analytical and numerical ways,
respectively.

\subsubsection{$\beta < 0$}

\begin{figure}[ht]
\begin{center}
\mbox{\epsfig{figure=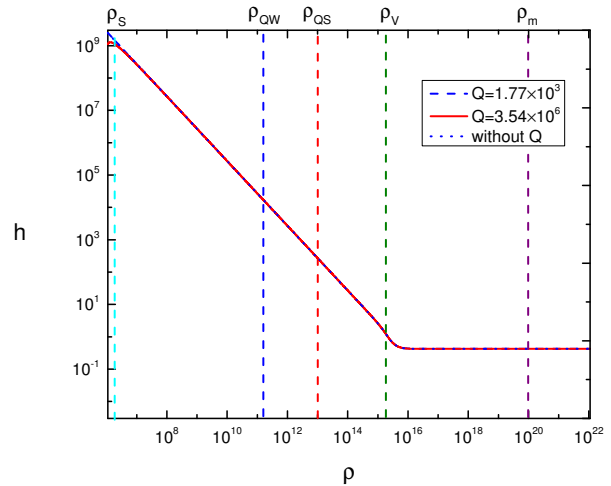,width=7.5cm,angle=-90}}
\caption{{\it
Plot of the evolutions of metric factor $h$ as functions of radial
coordinate $\rho$ in a charged solar system described by nonlinear massive
gravity. The parameters of the massive gravity model are taken as:
$\alpha=1$ and $\beta=-1/2$. Moreover, $m$ and $M$ are the same as those
provided in Fig \ref{Fig:Case2_h}.
}}
\label{Fig:Case3_h large}
\end{center}
\end{figure}

Let us first consider the subcase $\beta<0$. By solving $h$ numerically we
find that there is only one branch of solution which is real between the
Schwarzschild radius and infinity. It evolves from a very large value to a
small constant, as moving from the Schwarzschild radius to the Compton
wavelength. Around the Vainshtein radius it is of order ${\cal{O}}(1)$. A
strong charge suppresses the value of $h$ slightly when it is close to and
inside the Schwarzschild radius. Beyond this range the charge does not
cause an obvious effect. This can be seen in Fig. \ref{Fig:Case3_h large},
where the three curves almost coincide.

As shown in this figure  $h>>1$ in the range $r_S<\rho<\rho_Q$.
Thus, in the semi-analytical calculation, we keep the leading order terms
of the equations of motion for the metric factors, and then we acquire
\begin{eqnarray}
 h &\simeq&
\frac{1}{\beta^{1/3}}(\frac{\rho_Q{}^4}{\rho^4}-\frac{\rho_V{}^3}{\rho^3})^
{1/3},\\
 n' &\simeq& -\frac{m^2}{2\beta^{1/3}}
(\frac{\rho_Q{}^4}{\rho}-{\rho_V{}^3})^{1/3},\\
 \tilde{f} &\simeq& -\frac{G Q^2}{8\pi \rho^2} +\frac{\alpha m^2
}{2\beta^{2/3}}(\frac{\rho_Q{}^4}{\rho}-\rho_V{}^3)^{2/3}\nonumber\\
 &&-\frac{m^2}{2\beta^{1/3}} (\rho_Q{}^4-\rho \rho_V{}^3)^{1/3}\rho^{2/3}.
\end{eqnarray}
From the above semi-analytic results we find that the corrections to
$\tilde{f}$ and $n$ are so dramatic that the usual Schwarzschild-like
gravitational potential (in form of $1/\rho$) is exactly canceled. This
result agrees with the conclusion of \cite{Koyama:2011yg} in which a
neutral solar system was considered.

Finally, for the gravitational potential associated with the electric
charge, the RN-like factor $\frac{Q^2}{\rho^2}$ in $g_{tt}$ also
disappears, but the dominant term in the small radius regime roughly takes
the form of $\rho^{2/3}$. For the $g_{rr}$ component the sign in
front of $\frac{Q^2}{\rho^2}$ changes from positive to negative compared
to the General Relativity result. These new features suggest that the charged solar system
described by nonlinear massive gravity is much more smooth near the origin
than that described by General Relativity. However, since below the Vainshtein radius the
difference from General Relativity is quite significant in this case, the corresponding
parameter space is likely ruled out by local, solar
system experiments.

\subsubsection{$\beta > 0$}

\begin{figure}[ht]
\begin{center}
\mbox{\epsfig{figure=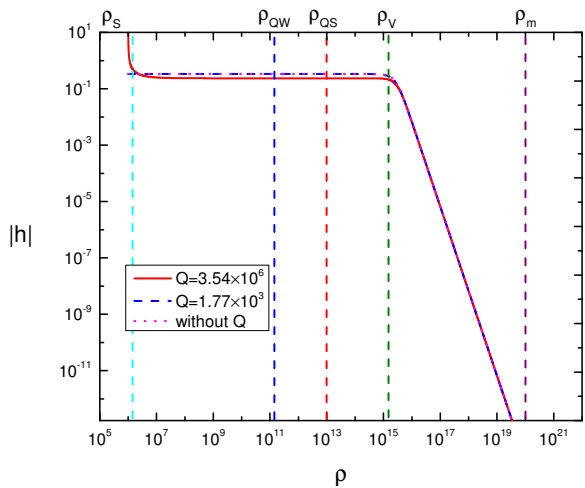,width=7.5cm,angle=-90}}
\caption{{\it
Plot of the evolutions of metric factor $h$ as functions of radial
coordinate $\rho$ in a charged solar system described by nonlinear massive
gravity. The parameters of the massive gravity model are taken as:
$\alpha=1$ and $\beta=3$. Other parameters are the same as those used in
Fig. \ref{Fig:Case3_h large}.
}}
\label{Fig:Case3_h small}
\end{center}
\end{figure}

\begin{figure}[ht]
\begin{center}
\mbox{\epsfig{figure=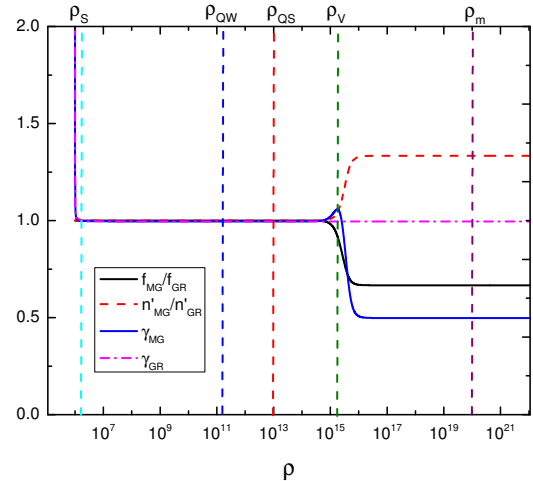,width=7.5cm,angle=-90}}
\caption{{\it
Plot of the evolutions of the ratios $n'/n_{GR}'$,
$\tilde{f}/\tilde{f}_{GR}$, and the quotient $\gamma \equiv \frac{\tilde
f'}{n'}$ as functions of radial coordinate $\rho$ in a charged solar system
described by nonlinear massive gravity. The parameters of the massive
gravity model are taken as: $\alpha=1$, $\beta=3$. Other parameters are the
same as those used in Fig. \ref{Fig:Case3_h large}. The `$f$' in the plot
represents the metric factor $\tilde{f}$ in the main text.
}}
\label{Fig:Case3_GR}
\end{center}
\end{figure}

We  now consider the subcase $\beta>0$. Apart from the previous solution in
which $h$
is large-valued below the Vainshtein radius, there exists a second solution
in which the metric factor $h$ always takes a small value outside the
Schwarzschild radius. Similarly to the previous subcase we solve
the equations of motion numerically, obtaining the solution for the metric
factor $h$ as shown in Fig. \ref{Fig:Case3_h small}.
In addition, we extract the evolutions of $\tilde{f}$ and $n'$ along the
radial coordinate $\rho$. In order to provide a clearer picture of the
difference of the dynamics in massive gravity from that of General Relativity, in Fig.
\ref{Fig:Case3_GR} we plot the ratios $n'/n_{GR}'$ and
$\tilde{f}/\tilde{f}_{GR}$, respectively.

From Fig. \ref{Fig:Case3_h small} we can see that in a strongly charged
solar system the absolute value $|h|$ can become larger than $1$ inside
the Schwarzschild radius. As we move away from the Schwarzschild radius
$h$ evolves to a constant, which coincides with the value when the charge
is weak. After crossing the Vainshtein radius $h$ approaches $0$ rapidly.
By comparing $\tilde{f}$ and $n'$ with the results of General Relativity, we can clearly
see the Vainshtein mechanism from Fig \ref{Fig:Case3_GR}.

Since $|h|$ is always smaller than unity outside the Schwarzschild radius,
we can solve its equation of motion by neglecting the terms proportional to
$h^3$ and $h^5$. Correspondingly, the approximate solution is given by
\begin{eqnarray}
 h &\simeq& \frac{3\rho^4-2\alpha
\rho_Q{}^4}{12(\alpha\rho^4-2\beta\rho_Q{}^4+2\beta\rho\rho_V{}^3)} \times
\nonumber\\
 && \left\{1-\left[{1+\frac{48\rho
\rho_V{}^3(\alpha\rho^4-2\beta\rho_Q{}^4+2\beta\rho\rho_V{}^3)}{
(3\rho^4-2\alpha\rho_Q{}^4)^2}}\right]^{1/2}\right\}
~,\nonumber\\
\end{eqnarray}
which proves to be in good agreement with the numerical solution. The
corresponding expressions for $\tilde f$ and $n'$ are obtained
from  equations \eqref{solf} and
\eqref{solnp}, however they are quite complicated and thus we do not
present them explicitly. Furthermore, we would like to point out that in
this case the dynamical features of the metric factors $\tilde{f}$ and $n$
are quite similar to the case of $\alpha\neq0$ and $\beta=0$, as can be
seen by comparing Fig. \ref{Fig:Case3_GR} and Fig. \ref{Fig:Case2_nf}. This
implies that the solution obtained in the $\beta=0$ case might be
dynamically stable in the parameter space of the nonlinear massive gravity
model.

\subsubsection{$\beta=\alpha^2/6$: An exact analytic solution}

In last subsection we examine a family of solutions in the model of
nonlinear massive gravity under the particular parameter choice
$\beta=\alpha^2/6$. This special parameter choice was first noticed in \cite{Nieuwenhuizen:2011sq}, where the exact SdS and RN-dS solutions were
obtained with an arbitrary cosmological constant term.
This relation was also applied by the authors of \cite{Berezhiani:2011mt},
who constructed a special family of black hole solutions on a fixed dS
background.

We insert the relation $\beta=\alpha^2/6$ into the nonlinear equation of
motion \eqref{solh} and we find that there exists a special solution for
$h$, namely
\begin{eqnarray}\label{h_exact}
 h=\frac{1}{\alpha}~,
\end{eqnarray}
which implies a constant metric factor $H=(1+\alpha)/\alpha$. Then we
substitute the solution \eqref{h_exact} into the exact background
equation of motion \eqref{EoM} and the Bianchi constraint \eqref{EoMBi}. It
is easy to verify that the constraint equation \eqref{EoMBi} is
automatically satisfied. Working with the $r$ coordinate directly we
see that the main equation \eqref{EoM} yields the following exact solution:
\begin{eqnarray}
 N^2(r) &=& F^2(r) \nonumber\\
 &=& \frac{(1+\alpha)^2}{\alpha^2} +
\frac{GQ^2(1+\alpha)^4}{4\pi\alpha^4r^2}-\frac{r_M}{r}-\frac{m^2r^2}{
3\alpha},\ \ \ \
\end{eqnarray}
where $r_M$ is an integration constant.

We can perform the following coordinate rescaling:
\begin{eqnarray}
 t \rightarrow \frac{\alpha}{1+\alpha}t~,~~r \rightarrow
\frac{1+\alpha}{\alpha}r~,
\end{eqnarray}
and introduce two coefficients
\begin{eqnarray}\label{r_Lambda}
 \tilde{r}_S \equiv \frac{\alpha^3r_M}{(1+\alpha)^3}~,~~r_\Lambda \equiv
\frac{\sqrt{3\alpha}}{m}~,
\end{eqnarray}
which are related to the Schwarzschild radius and the de Sitter radius,
respectively. We get the exact form of the RN-dS-like solution
as
\begin{eqnarray}
 ds^2 = -A(r)dt^2+\frac{dr^2}{A(r)}+r^2d\Omega^2~,
\end{eqnarray}
with
\begin{eqnarray}
 A(r) = 1+\frac{r_Q^2}{r^2}-\frac{\tilde{r}_S}{r}-\frac{r^2}{r_\Lambda^2}~,
\end{eqnarray}
where the forms of $r_Q$, $\tilde{r}_S$ and $r_\Lambda$ are provided in
\eqref{r_Q} and \eqref{r_Lambda}, respectively. The above
solution can recover the standard RN result in  General Relativity, and
$\tilde{r}_S$ coincides to the usual Schwarzschild radius $r_S$ when we
 take $m=0$. Furthermore, our result is in agreement with the one
obtained in  \cite{Nieuwenhuizen:2011sq}, however we did not introduce an
additional cosmological constant in order to see whether and how a pure
massive gravity model can yield a dS background by itself.

Moreover, the coefficient $r_\Lambda$ is determined by the combination of
the graviton mass $m$ and the model parameter $\alpha$. When $\alpha$ is of
order ${\cal{O}}(1)$ we can apparently observe that the property of dS
background is completely determined by the graviton mass. If we apply this
feature in a cosmological setup, the existence of a tiny graviton mass can
drive a late-time acceleration of the universe and thus might explain
the present cosmological observations. However, if we tune $\alpha$ to an
extremely large value then the effect of the graviton mass can be decreased
and the background dynamics approach those of General Relativity. Mathematically
this effect can be seen by the way ${\cal K}$ appearing in the graviton
potential ${\cal U}$ is roughly proportional to $1/\alpha$ and thus it
yields the amplitude of the effective energy-momentum tensor $X_{\mu\nu}$
in form of $1/\alpha$. Eventually, the behavior of nonlinear massive
gravity approaches standard Einstein gravity in the limit of
$\alpha\gg1$.

\section{Conclusions}
\label{Conclusions}

In the present work, we investigated the spherically symmetric solutions of
a charged solar system in the context of nonlinear massive gravity. Due to properties 
of the graviton potential in the dRGT model,
the BD ghost which historically plagues all massive gravity theories can be
removed. However, inherited from other massive gravity models, the
longitudinal mode of gravitons is strongly coupled and thus greatly affect
the gravitational potential at macroscopic scales. Therefore, this model is
expected to be constrained by solar system observations.

Depending on the different dynamics of our solutions, the solution parameter
space can be roughly categorized into three parts:
\begin{itemize}
\item{The first class corresponds to $\alpha=\beta=0$ and thus the
graviton's potential takes a fixed form. The solution in this subclass is
well described at the perturbative level, but the vDVZ discontinuity cannot be
avoided. However, the post-Newtonian parameter in this class fails to agree
with General Relativity and thus the corresponding parameter choice is observationally
ruled out.}
\item{In the second subclass, we keep $\beta=0$ but we allow $\alpha$ to be
an arbitrary constant. The corresponding solution shows that General Relativity
 can be recovered between the outer horizon of the black hole and
the Vainshtein radius by virtue of the Vainshtein mechanism. This scenario
is similar to the case of the neutral black hole in massive gravity. However,
the existence of an electric charge could increase the value of
the metric factor $h$ within a newly defined radius $\rho_Q$ and thus the
detailed evolutions of time-like and space-like metric components behave
differently from those of a neutral black hole. Namely, the metric factor
$n$ obtains a logarithmic correction when the radius is close to the outer
horizon.}
\item{The third subclass of parameter choice is the most general in the
 parameter space, which requires both $\alpha$ and $\beta$ to be
non-vanishing. In this case the dynamics of solutions behave dramatically
different depending on the positivity of $\beta$. When $\beta<0$ the
strongly coupled scalar graviton greatly decreases the strength of gravity
at small length scales, and thus the usual Schwarzschild-like gravitational
potential totally disappears which severely challenges all astronomical
observations. However, if $\beta$ is positive General Relativity  can be
recovered again through the Vainshtein mechanism. This behavior is similar
to the solution in the second subclass with $\beta=0$. Therefore, the
solution in this case, together with the solution in the second subclass,
might provide a certain parameter space for nonlinear massive gravity to
conform with current solar system observations.}
\item{Finally, there exists a particular parameter choice in the last
subclass which suggests $\beta=\alpha^2/6$. Under this condition the
background equations of motion can be solved exactly and yield a solution
which is identical to the RN-dS form in which only the dS radius
$r_\Lambda$ contains the model parameter $\alpha$. The exact solution with
such a special parameter choice can recover the standard result in General Relativity in the
limit of either a vanishing graviton mass or an extremely large value of
the parameter $\alpha$.}
\end{itemize}

Note that, the structure of solutions does not only depend on the sign of
$\beta$, but also on a diverse structure in the parameter space of $\alpha$ and
$\beta$. The authors of \cite{Sbisa:2012zk} analyzed the whole parameter
space of a neutral solar system under nonlinear massive gravity and found
that each region showed a completely different behavior with respect to the
inner solutions (near the body) and the asymptotic solutions, and thus
affect the Vainshtein mechanism. It would be interesting to perform a
global analysis on the parameter space in our case too.

Lastly, we would like to mention that in the present work we only focus
on the analysis of background solutions of a charged solar system in the
dRGT model, without studying the perturbations. At the background level
the parameter space of model parameter has already shown rich
behaviors and a sizable regime has already been ruled out by
observations. Due to the Vainshtein mechanism there exists an island in the
 parameter space which is consistent with astronomical data at the
background level. Moreover, this situation could  dramatically changed if
the perturbations were taken into account and we expect the model
parameters appearing in the nonlinear massive gravity would be further
constrained. However, we leave such a study for future investigation.

\begin{acknowledgments}
It is a pleasure to thank Lasha Berezhiani, Giga Chkareuli, Claudia de Rham,
Gregory Gabadadze, Kazuya Koyama, Gustavo Niz, Gianmassimo Tasinato, and
Andrew Tolley for useful comments on the manuscript. The work of CYF and
DAE is supported in part by the DOE under DE-SC0008016 and the Cosmology Initiative at Arizona State
University. The work of GCX is supported in part by the Summer Research
Assistantship from the Graduate School of the University of Mississippi and
in part by the Cosmology Initiative in Arizona State University. The
research of ENS is implemented within the framework of the Action
Supporting Postdoctoral Researchers of the Operational Program ``Education
and Lifelong Learning'' (Actions Beneficiary: General Secretariat for
Research and Technology), and is co-financed by the European Social Fund
(ESF) and the Greek State.
\end{acknowledgments}

\end{document}